\documentclass{scrartcl}
\usepackage[utf8]{inputenc}
\usepackage[T1]{fontenc}
\usepackage[british]{babel}
\usepackage{graphicx}
\usepackage{amsmath}
\usepackage{libertinus}
\usepackage{xurl} 
\usepackage[hidelinks]{hyperref}
\usepackage{changepage}
\usepackage{diagbox}
\usepackage{xcolor}
\usepackage{wasysym}
\usepackage{tikzsymbols}
\usepackage[round]{natbib}
\setcitestyle{aysep={}}
\usepackage{enumitem}
\ifdefined\LaTeXML
  \author{Fredrik Jansson\\[0.6em]
    \textit{Department of Business and Mathematics, M\"alardalen University, V\"aster\aa s, Sweden}\\
    \textit{Centre for Cultural Evolution, Department of Psychology, Stockholm University, Sweden}\\
    \textit{Institute for Futures Studies, Stockholm, Sweden}}
  \date{Preprint\footnote{Chapter in Johan Lind and Anna Jon-And, eds. (in press): \emph{Cultural Evolution from Minimal Principles}. Cambridge, United Kingdom: Cambridge University Press.}\footnote{This work was supported by the Marianne and Marcus Wallenberg Foundation (2021.0039).}}

\else
  \usepackage{authblk}
  \author{Fredrik Jansson}
  \affil{\vspace{1em}Department of Business and Mathematics, M\"alardalen University, V\"aster\aa s, Sweden}
  \affil{Centre for Cultural Evolution, Department of Psychology, Stockholm University, Sweden}
  \affil{Institute for Futures Studies, Stockholm, Sweden}
  \date{Preprint\footnote{Chapter in Johan Lind and Anna Jon-And, eds. (in press): \emph{Cultural Evolution from Minimal Principles}. Cambridge, United Kingdom: Cambridge University Press.} \footnote{This work was supported by the Marianne and Marcus Wallenberg Foundation (2021.0039).}\\[-1em]}
\fi

\AtBeginEnvironment{thebibliography}{\raggedright}
\setlist[itemize]{noitemsep}
\setlist[enumerate]{noitemsep}
\ifdefined\LaTeXML
  \providecommand{\setcapindent}[1]{}
  \providecommand{\setkomafont}[2]{}
  \providecommand{\addtokomafont}[2]{}
\fi
\setcapindent{0pt}
\setkomafont{captionlabel}{\bfseries}

\title{Modelling cultural evolution}

\begin{document}
\maketitle

\begin{abstract}
    \noindent Formal modelling provides a toolkit for understanding cultural dynamics, from individual decisions to recurring patterns of change. This chapter explains what models are and why they matter. Using a precise, shared language, they aid thinking and communication by turning fuzzy assumptions into clear, comparable, testable claims. The chapter describes the modelling process, trading explanatory clarity against predictive specificity. Four families of models are surveyed, from the micro-level with optimising agents to macro-level dynamics with heuristic or even implicit agents, covering reasoning (Bayesian inference, game theory), adaptive updating (reinforcement learning, evolutionary games), mean-field approaches (compartmental models, population dynamics), and complex systems (agent-based models, social networks). Building on these, a general template for modelling cultural evolution is outlined that connects system states, cognitive processes, behaviour, and macro-level outcomes in dynamic loops, linking individuals, groups, institutions, and their environments. Taken together, these tools support a pluralist but coherent understanding of cultural change.

\end{abstract}

\section{Introduction}\label{introduction}

Imagine you are on a trip with someone special and would like to spend some quality time together. You are looking for a calm restaurant with good cooking. You pass by several restaurants that look decent and really calm, until you end up queueing for the busiest of them. What happened? Well, the first restaurants were calm because they had almost no guests, and this one must be more a popular choice for a reason. Even if all the other guests also look like tourists, they seem to know something that you do not. This place probably has the best food, and the other ones are likely of low quality, so why risk it? It is worth the wait.

To describe this situation in slightly more formal terms, you observed a heavily skewed distribution of guests over the restaurants. You quickly made a speculation of the process that led to this distribution: some restaurants serve better food than others, and those restaurants attract more customers, while those with few guests have repulsed customers due to bad food or service, or high prices. Perhaps unconsciously, you have made a mental model of the situation and used it to make an informed decision.

A mental model is an internal representation of an external reality. But it needs not stay at that: you might make a case to your partner why you think you should spend half an hour waiting for a table at a place where you will hardly be able to hear each other. Designing and presenting abstract models may or may not sound like a romantic activity to start the evening, but it is a fairly accurate depiction of what happens next, and something that people do all the time. When communicating your understanding of the situation, you are building a coarsely simplified version of it. There is no need to account for the moustache of the person behind you in the line, the shape of the cobbled stone you are standing on, or even the time of day; you just focus on the key parts: good restaurants attract many people, poor ones do not; there is a crowd queuing for this restaurant, the one next door is empty; ergo, this is the better restaurant. You have designed an abstract model: a structure that can potentially represent real-world phenomena. You will probably stay at that, letting the verbal model and the intuitive reasoning suffice to make you satisfied with your choice, and move on to talking about other topics.

Luckily, we here have the space to elaborate on this and develop a bit more formal a model, and look at the situation from the outside. We will search for the simplest explanation that can describe the phenomenon. For this reason, with our outsider's eyes, we can note that there are at least two processes going on in this story: people have knowledge of the quality of restaurants and people are copying the behaviours of others. The latter is a crucial part of the story, at least for the romantic couple who do not possess the requisite knowledge. Let us see how far we can get then by assuming only the latter process.

\begin{figure}[tbp]
\centering
\includegraphics[width=1\linewidth]{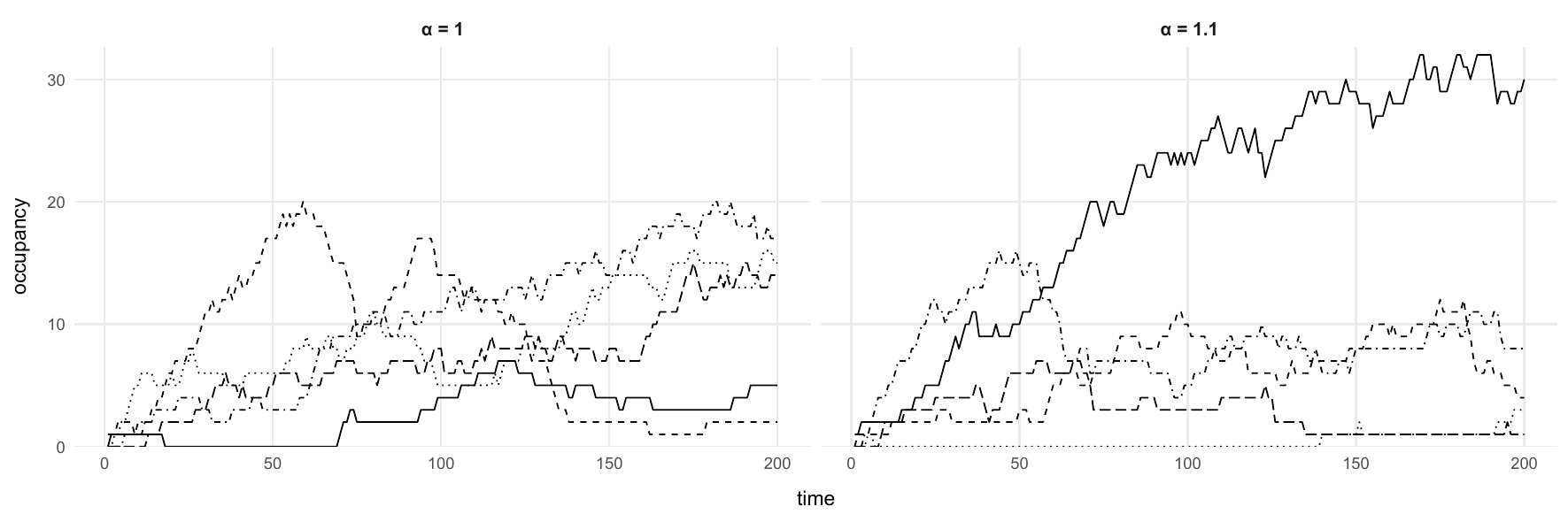}
\caption{Simulations without and with a conformity bias. The number of guests at each of five restaurants is plotted over time.}
\label{fig1}
\end{figure}

Let us assume then that you pick a person uniformly at random and copy whatever choice that person made. If this sounds unrealistic and far off what happens in the situation with the restaurants, then it can be equivalently formulated as choosing restaurant with a probability proportionate to the number of current guests. We can start by assuming that it is not only proportionate but also equal, that is,

\[p_{i} = \frac{n_{i} + 1}{{(n}_{1} + 1) + {(n}_{2} + 1) + \cdots + {(n}_{k} + 1)}
\]
if $p_i$ is the probability of choosing restaurant $i$ and $n_i$ is the number of guests at restaurant $i$. One is added to each restaurant to give them a nonzero probability to be chosen. We iterate this over discrete time steps, such that at each time step, there is one new guest making a decision according to the specified probability, and each guest at all restaurants leaves with a certain probability, say 2\%. A simulation of this model with five restaurants is presented in Figure~\ref{fig1}a. We can see that the popularity of restaurants is already unevenly distributed given these simple assumptions. However, we can also note that there are no self-reinforcing effects; which restaurant is the most popular one even varies over the evening. As expected, we need to develop the model further.

Before considering the more complex elements of food quality, reputation and previous experience, however, we can try to tweak the assumption on how social copying takes place, informed by cultural evolution theory. It has been argued \citep[e.g.]{henrich1998evolution} that people often employ a conformity bias, meaning that we copy common behaviours disproportionately more often than rare ones. In our model, this could be implemented by raising the number of guests to a power \(\alpha\), that is,

\[p_{i} = \frac{\left( n_{i} + 1 \right)^{\alpha}}{{{(n}_{1} + 1)}^{\alpha} + {{(n}_{2} + 1)}^{\alpha} + \cdots + {{(n}_{k} + 1)}^{\alpha}}
\]

This example hopefully illustrates the high prevalence of models, the modelling process and the use of formalising models. Models are everywhere: we make our own mental models and use them to make sense of the world, and then we turn them into verbal models to communicate them to others. Our brains are limited, so it is a physical necessity that reality must be simplified. Leaving the everyday situation and moving into scientific territory, verbal models can be turned into theories and hypotheses, and formalised for systematic scrutiny. In this chapter, we will focus on the formal part of modelling cultural evolutionary processes, but the models typically originate in everyday observations. As a striking example, with a comparison from the natural sciences, Newton's law of gravity famously originated from everyday observation: as tradition recounts, an apple fell from a tree, prompting Isaac Newton to ponder why objects fall straight towards the earth. This simple observation led him to hypothesise a universal force of attraction between masses \citep{westfall1980never}. Newton formalised this intuition into the mathematical law of universal gravitation, a model describing how any two objects attract each other with a force proportional to their masses and inversely proportional to the square of the distance between their centres \citep{newton1687philosophiae}.

Apples falling to the ground and planetary orbits are disparate but highly regular patterns of change that could both be explained by a simple mechanism of objects attracting each other. Cultural and societal change is clearly not as regular, but it is still often possible to discern general empirical patterns. To start with, culture might not change much at all. For instance, while languages evolve, fundamental aspects like grammar, and the number of words in a language, can exhibit significant persistence \citep{pagel2007frequency}. Within the same cultural domain, such as moral opinions, there is sometimes little change, such as public opinion in the US on abortion rights or prohibition of pornography, while the general acceptance of gay marriage has increased linearly over the past 50 years \citep{mulligan2013dynamics,strimling2019connection}. Cultural evolution can also exhibit rapid, accelerating growth in certain domains. Some examples are the number of publications in a scientific field \citep{bornmann2015growth}, retweets or the viral spread of information on social media \citep{vosoughi2018spread}, and the number of transistors on a microchip (Moore's law) \citep{schaller1997moores}. Sometimes growth is limited and exhibits logistic s-shaped curves \citep{rogers1962diffusion}, such as the propagation of innovations in a population, for example mobile subscriptions \citep{gruber2001evolution}. For some phenomena, there are clear tipping points with threshold effects, where a period of little change is followed by an outburst in dissemination, such as was the case with the \#metoo movement \citep{centola2018experimental} or the Arab spring \citep{howard2011opening}. Other phenomena work like fashion trends, with their adoption following an inverted u-shape \citep{acerbi2012logic}. These inverted u-shapes can recur periodically, working like trend cycles, such as business cycles \citep{burns1946measuring}. To understand why these and other patterns recur across cultural domains, and why certain patterns are matched to certain phenomena, we would like to identify reasonable processes that generate these patterns.

\begin{figure}[tbp]
\centering
\includegraphics[width=1\linewidth]{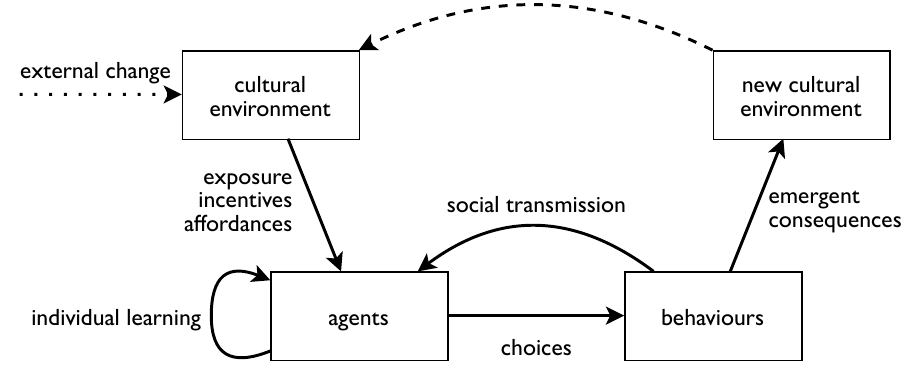}
\caption{Adaptation of Coleman’s boat for cultural evolution. Relationship between macrolevel cultural context and microlevel individual actions.}
\label{fig2}
\end{figure}

The described patterns of change are all observed at the population level, or macrolevel. However, these patterns are not generated through purely populational processes, but they all involve the actions of individual agents, at the microlevel. The interplay between the macrolevel and the microlevel can be depicted through a diagram (see Figure~\ref{fig2}) often referred to as Coleman's boat \citep{coleman1973mathematics}: changes at the macrolevel affect agents at the microlevel, whose behaviours consequently change, further influencing other agents through cultural transmission. The behavioural changes among agents amass into consequences at the macrolevel, often not as pure aggregations, but as emergent patterns that occur only through the interaction of several agents. These consequences then bring with them changes at the macrolevel, and the cycle starts anew. Our initial model on choosing restaurants through proportionate copying assumed only the trivial interaction of pure aggregations, but when we introduced a conformity bias, a macrolevel change in the number of guests affected agents disproportionately, which combined into a rich-get-richer effect for the restaurants. A more intriguing example that draws on more social theory is how improved social conditions can lead to revolution by first inducing frustration among individuals whose expectations are raised but not fulfilled, leading to aggression, which spreads frustration to other individuals, and ultimately, this collective aggression culminates in revolution when a critical amount of people are engaged \citep{coleman1990foundations}.

The complexity of these micro--macro interactions quickly surpasses our cognitive limits, making informal or purely verbal reasoning insufficient. Formal modelling is a highly useful tool here: it compels us to clarify our assumptions explicitly, reveals logical consequences that might otherwise remain hidden, and allows systematic exploration of how individual-level mechanisms scale up into macrolevel patterns. By employing formal models, we can rigorously test competing hypotheses, quantify the relative importance of different microlevel processes, and identify thresholds or tipping points that verbal explanations alone often obscure. Ultimately, formal models offer a precise language to understand cultural evolution, connecting theoretical insights closely with empirical data.

\section{On modelling}\label{on-modelling}

To make the concept of a formal model a bit more formal (pun intended), we can define it as a simplified version of a system with a specification of parts and relationships. The simplification is not only a necessity, but also a main feature. To find your way from A to B, a map with the right amount of granularity is more helpful than an overly detailed one. The right amount of granularity depends on your aims: just compare a tourist map for walking around the city centre to the one you find in the metro. You do not need to know the terrain, street names or locations of bridges to find out which trains will take you to a certain metro station. While maps describe a static structure, understanding cultural evolution typically entails describing a process; we are thus interested in dynamic models. If we manage to describe them using the language of mathematics, then we have made them unambiguous, and also built a logical engine for turning assumptions into conclusions. This logical engine can then be used to test hypotheses formally.

Suppose you have a hypothesis A for what mechanism generates an observed phenomenon B. You design a formal model to investigate the consequences of A, and the output of the model is indeed B. What can you conclude? Well, you have not proven that A is the cause of B in the real world, because there may be other mechanisms that could also generate B. However, you have shown that A is \emph{sufficient} to generate B. This can be very informative, because it challenges more complex explanations. In our restaurant example, we saw that people having knowledge of what are good restaurants, homogeneous preferences and making informed choices are not necessary assumptions for generating highly skewed occupancy with a dominating restaurant; it is sufficient to assume a conformity bias. We also saw that assuming unbiased random copying (A) did not produce such high skewness (B), that is, A did not generate B. That does not necessarily mean that A is wrong, but it is insufficient and further, or possibly other, assumptions are necessary. In our more general model for choosing restaurants containing the parameter \(\alpha\), we found that it is \emph{necessary} to assume that \(\alpha > 1\), that is, some kind of bias, at least when keeping the other parameters constant.

Finally, a model might show that A leads to C instead of B. If we happen to know that C is false, then we can deduce the powerful conclusion that A is also false (an inferential rule called \emph{modus tollens} or \emph{denying the consequent}). If we had reason to believe A in the first place, then falsifying A is real scientific progress in the Popperian sense \citep{popper1962conjectures}. This has been the destiny of, for example, Ptolemaic astronomy, placing earth at the centre of universe with planets and stars moving around it in perfect circles, producing predictions inconsistent with observed planetary motion, such as Mars' retrograde motion \citep{kuhn1957copernican,kuhn1962structure}. As another example, from cultural evolution, it is commonly assumed that rational choice guides people's decisions. Applied to a cultural market such as pop charts, people should choose songs based on their quality rather than peer influence. Comparing conditions in an experiment where respondents could listen to new music with or without social information about other users' choices, it was shown that lack of social information produces more evenly distributed success rates while also making success of individual songs more predictable \citep{salganik2006experimental}. In contrast, in reality pop album charts exhibit power law distributions where it is hard to predict individual success -- a pattern that is generated by just assuming random copying \citep{bentley2007regular,eriksson2010bentleys}. In a market with social information, random copying is arguably a null hypothesis, a model that can serve as a reference when we add further assumptions. Adding assumptions on informed rational choices in this case leads to worse predictions.

The descriptions of what formal models are and what they do has hopefully also served to show why they are useful. If we expand on the latter a bit, we might summarise the reasons for mathematical modelling in four key properties: precision, prediction, tractability and thinking \citep[see][]{smaldino2023modeling}.

Precision means clearly articulating what is and is not included in the model. By explicitly defining assumptions, relationships, and constraints, formal models minimise ambiguity, avoid hidden assumptions and enable precise communication. Because mathematical models rely on a common symbolic language, different researchers can understand, replicate, and extend each other's work. This is particularly useful when working across disciplines, where people might use different terms to describe the same structures.

Prediction involves uncovering consequences of explicitly stated assumptions. Once the model's premises are defined, the logical framework makes it possible to predict outcomes that must follow from these assumptions. This makes the scope of applicability transparent, highlighting clearly what phenomena the model can reliably explain and under what conditions. Indirectly, this enables hypothesis testing, because clear predictions can be falsified, but often the predictions themselves can be the target of the model. The weather forecast is an example where prediction is the key focus of the model.

Tractability refers to the model's analytical manageability, which due to its simplifications and precision allows for formal exploration and clear demonstrations of how specific mechanisms generate observed patterns. With clearly defined components and relationships, we can rigorously demonstrate how certain mechanisms generate observed patterns. As discussed above, we can test whether certain hypotheses are sufficient (capable of producing a result), necessary (indispensable to producing a result), or even false (producing incorrect predictions). This formal clarity identifies which assumptions are essential and which are misleading.

Finally, mathematical models serve as mental frameworks that help us think about complex dynamics in structured ways. Models serve as concrete analogies or mental tools for understanding entangled interactions, helping us grasp intricate dynamics through simpler, abstract representations. This is related to the use of metaphors, which also function as abstract analogies to facilitate thinking and allow us to identify and categorise different types of situations more intuitively. By formalising metaphors into models, we can test our intuition. Often, mathematical models allow us to explore more complex interactions than unaided intuition can do.

While all these four properties are reasons for modelling, we may put different emphasis on them depending on our objectives: do we want to gain an understanding of an observed phenomenon, do we want to find a general mechanism, or do we rather want to make accurate predictions into the future? How many of the complexities of a system do we need to capture? We can make a map of the types of models to aim for depending on our approach and level of detail \citep{smaldino2020how}.

The approach ranges from analytical to computational models. Analytical models are typically described by equations that can be analysed and solved, meaning that we know the exact values of all variables at any time step and the relative contribution from each. We can thus gain a full understanding of the system. This type of analysis is possible also with some interactions between entities in dynamical systems, but with more entities and interactions, the model quickly becomes analytically intractable. Often we can find approximate numerical solutions, or in the more complex cases, simulate the system behaviour over time. One simplifying assumption that can enable a model to be written down as a simple equation system is that the agents are homogeneous. In our model of restaurant choices, all agents follow the same rules, albeit stochastically. In reality, people have different knowledge and experiences, which can make them employ different rules, and even update them given new experience as the model unfolds. This heterogeneity can be captured by agent-based models (ABMs), which are computer programs simulating the actions and interactions of autonomous agents. ABMs are more flexible and can potentially be more realistic than equation-based models, but come at the expense of decreased tractability and understanding and a need for more data to support the increased amount of assumptions that follow with more degrees of freedom.

Models can describe phenomena at various levels of detail. Ideally, we might prefer highly detailed models capable of making precise, high-resolution predictions. However, when modelling complex systems such as human behaviour and societal processes, achieving high resolution necessitates incorporating numerous variables and assumptions. Each additional parameter increases the complexity and demands extensive data to accurately calibrate the model and validate underlying assumptions. Furthermore, many human behaviours and societal phenomena can be operationalised in multiple ways, and selecting appropriate measures is challenging. A data-driven model presumes valid operationalisations, with a clear correspondence between the parameter in the model and the empirically measured variable. For example, consider modelling voting behaviour in elections. A highly detailed model might include variables such as individual psychological traits, socioeconomic status, peer influence, media exposure, historical voting patterns, and regional cultural factors. Each variable can be operationalised differently; for instance, socioeconomic status might be measured by income, education level, occupation, or a combination of these. The different operationalisations might interact differently with the various operationalisations of the other variables. Adding parameters requires specific empirical validation, and the abundance of potential parameters increases the risk of misalignment between model parameters and real-world variables. Achieving high resolution is sometimes possible, for example in simpler systems, such as in physics, or in specific cultural domains with large amounts of data and clear operationalisations. For instance, in online digital media, user behaviour can often be precisely tracked through clicks, views, shares, and likes, with clear operationalisation and measurement, allowing calibration of predictive models. This is used for example to predict user engagement, personalise content recommendations and optimise advertising strategies. Typically, however, coarse-grained predictions are our best hope, especially if we also want to keep models tractable. Many models are therefore more conceptual than truly predictive, identifying pattern-generating mechanisms.

\begin{figure}[tbp]
\centering
\includegraphics[width=1\linewidth]{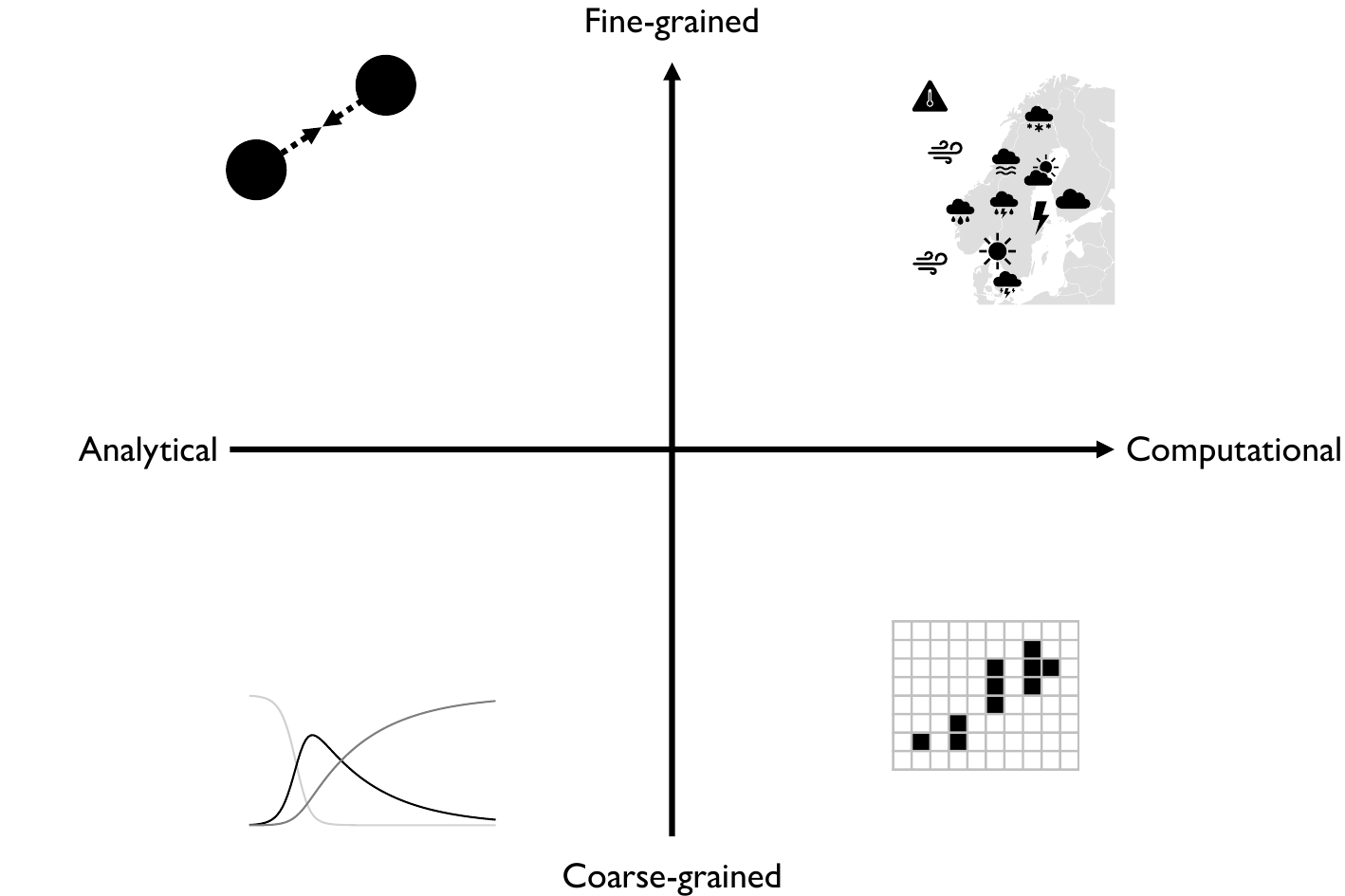}
\caption{Modelling approaches in two dimensions.}
\label{fig3}
\end{figure}

Combining the two dimensions for types of models, we get a map with four corners (see Figure~\ref{fig3}). In one corner we find the desirable analytical and fine-grained models, more common when studying the physical than the social world, for example Newton's law of gravity. Moving into complex systems, such as the weather, or perhaps the example of specific behaviour on digital platforms, we can still retain a high level of detail but only with a computational approach. If we instead want to retain the tractability of models, coarse-grained predictions may have to do, as found for example in epidemiology and SIR (tracing number of susceptible, infected and recovered agents) models. Many of us probably remember Covid pandemic strategies predicting the pattern of the spread of the contagion, rather than accurate forecasts of the number of infected people or in particular who would be infected. Models were used to coarsely evaluate the effectiveness of countermeasures to ``flatten the curve'' \citep[e.g.][]{ferguson2020report}. Finally, if we want to model complex behaviour, with homogeneous agents or many interacting components, say, then computational coarse-grained models may have to suffice, and may still be useful for providing general insights into system dynamics, evaluating different scenarios, guiding policy decisions, or identifying potential tipping points within the system. Agent-based models are typically found in this corner, and are employed frequently due to their versatility.

For all these types of models, there are some common properties and practices. Seeing a model as a system, we need to articulate the system, that is, decide its components and which components are interconnected. We further make explicit assumptions on the inner workings of the system and how it is influenced by factors external to the system, which we could call the environment. Thus, we need a representation of the internal states of the components, quantify the influence of the environment and decide on a structure of component behaviours and interactions. A simple example is the Lotka--Volterra predator--prey model \citep{lotka1925elements,volterra1926variazioni}. There are two components, representing two populations of prey (e.g. rabbits) and predators (e.g. foxes). The internal state of each component is its population size at a given time. As a result of environmental factors, the prey population grows naturally in the absence of predators and the predator population declines naturally due to mortality in the absence of prey. The two components interact in that the prey population decreases and the predator population increases proportionally to the number of predator-prey encounters (in turn assumed to take place proportionally to the sizes of the populations). This simple model generates two important outcomes: the sizes of the two populations tend to oscillate, with potentially large fluctuations, and, perhaps counterintuitively, making the environment better for the prey, increasing its growth rate, benefits the predator rather than the prey. Developing conceptual mathematical models like these are an exercise of the abductive principle, a formalised inference to the best explanation \citep{douven2021abduction}. No matter whether our goal is to predict with high precision or explain concepts, we can apply Occam's razor: make it as simple as possible, but as complicated as needed \citep{baker2022simplicity}. A common aim for models of cultural evolution is that they should be ``simple enough to be completely grasped and at the same time approximate how the world actually does work'' \citep{boyd1985culture}.

\begin{figure}[tbp]
\centering
\includegraphics[width=1\linewidth]{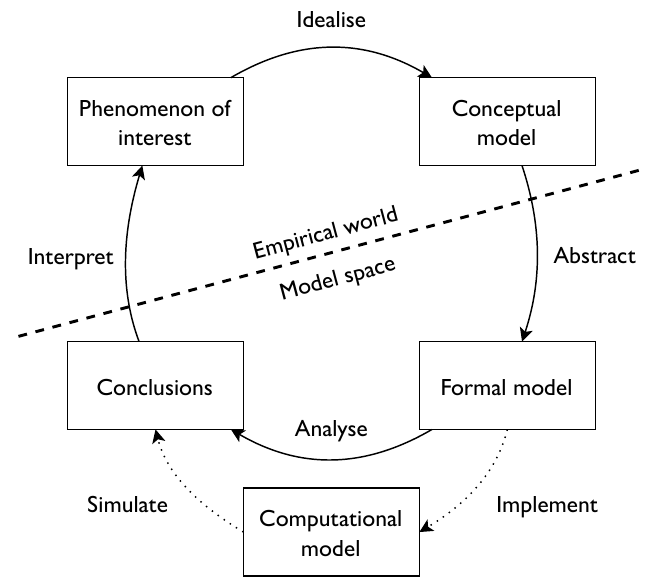}
\caption{A model of the modelling process.}
\label{fig4}
\end{figure}

We can summarise the modelling procedure in a model of mathematical modelling (see Figure~\ref{fig4}). We typically begin with a real-world problem, something concrete and observable that demands understanding or a solution. For instance, it could be about understanding why certain cultural phenomena, like the skewed popularity of restaurants, emerge and persist. To address this, we simplify the real-world scenario into a conceptual verbal model, focusing only on essential features while discarding irrelevant details. From this simplified scenario, we formalise and introduce abstract concepts, transforming the real model into a more precise and unambiguous mathematical model, defined explicitly through variables and equations. This mathematical form allows us to clearly articulate assumptions and relationships and serves as the basis for computation. Depending on its tractability, we can either draw conclusions directly from an analytical treatment, or if it is too complex, we translate the mathematical model into a computer model through programming, allowing us to perform calculations and simulate outcomes that would be challenging to compute analytically. Through analyses or simulations, then, we generate predictions and formal conclusions. We interpret these conclusions and relate them back to the real-world problem to assess their validity. The validity is often measured as the match between the model's predictions and empirical observations. Typically, there will be a considerable discrepancy after one iteration, so the cycle continues iteratively, refining each step to enhance the model's accuracy and explanatory power. As with all models, this model of the modelling process is a simplification, but hopefully it is accurate enough to be useful.

\section{Modelling paradigms in cultural evolution}\label{modelling-paradigms-in-cultural-evolution}

When modelling human actors, the components and interactions easily become complex in a sense that puts them in the computational and coarse-grained corner. However, this is not always necessary; depending on what phenomena we want to capture, sometimes simple models will still do. For example, many features of the aggregated motion of crowds of pedestrians, such as the development of lanes, oscillatory changes of walking directions, or jams in panic situations, can be captured by equation-based models from physics, even assuming that people are represented as particles \citep{helbing1995social,helbing2000simulating}. It will come as no surprise, though, that people cannot always be represented as particles: in contrast to physical inanimate objects, we are diverse; we possess agency in a way that makes us not only react to the environment; we are purposive, though error-prone, so even if internal or external forces guide us in particular ways, we may fail to follow them through; and we are both learning and socially influenced, so the environment and other components (a.k.a. people) of the system may change our internal states and how the different components interact. And as if social systems were not complex enough, also the dynamics of a single individual is already a complex outcome of many psychological and physiological processes. Importantly, though, with the example of behaviours of crowds in mind, or people choosing restaurants, qualitative properties of large-scale phenomena do not always depend on microscopic details.

Modelling human actors typically entails some representation of cognition. Depending on the phenomena we want to capture and their circumstances, it may suffice to assume that we are simply rule-based or that we are highly rational and thinking to solve problems optimally. The first assumption might apply in mindless low-stakes environments, such as movement in crowds, where people are happy if they manage just to move from A to B without bumping into other people. The second assumption, in contrast, might apply to high-stakes environments, for example when buying a house or when enterprises develop their corporate strategies. Usually, the answer lies somewhere in between, with adaptive agents following adaptive rules that are updated when the current strategy fails or when other agents following other strategies seem to do better.

Many different assumptions can make sense and produce useful models. There is no unifying model that rules them all, but we need many models to study different aspects of a phenomenon \citep{page2018model,smaldino2017models}. The parable of the blind men and an elephant serves as an analogy here: a group of blind men who have never come across an elephant before try to build a comprehension of what it is by touching one. They all touch one part each, and they all assert their own version of what an elephant is like: a snake, fan, tree or wall depending on whether they touch the tail, ear, leg or side. While their apprehensions are highly inconsistent, no one is wrong about the part they are studying. At the same time, the parable underscores the importance of seeking integrative or unifying theories. Though each blind man experiences only one aspect of the elephant, the elephant itself is real, coherent and unified. By zooming out, we lose detail, but we may be able to perceive the full elephant and synthesise the discoveries of the blind men. The zoomed-out perspective does not replace detailed scrutiny, but it can provide guidance to it. Interpreting the tail of an elephant benefits from having spotted the elephant. Cultural evolution theory might function as an elephant spotter, but it incorporates a wide range of models to study the different parts.

\begin{figure}[tbp]
\centering
\includegraphics[width=1\linewidth]{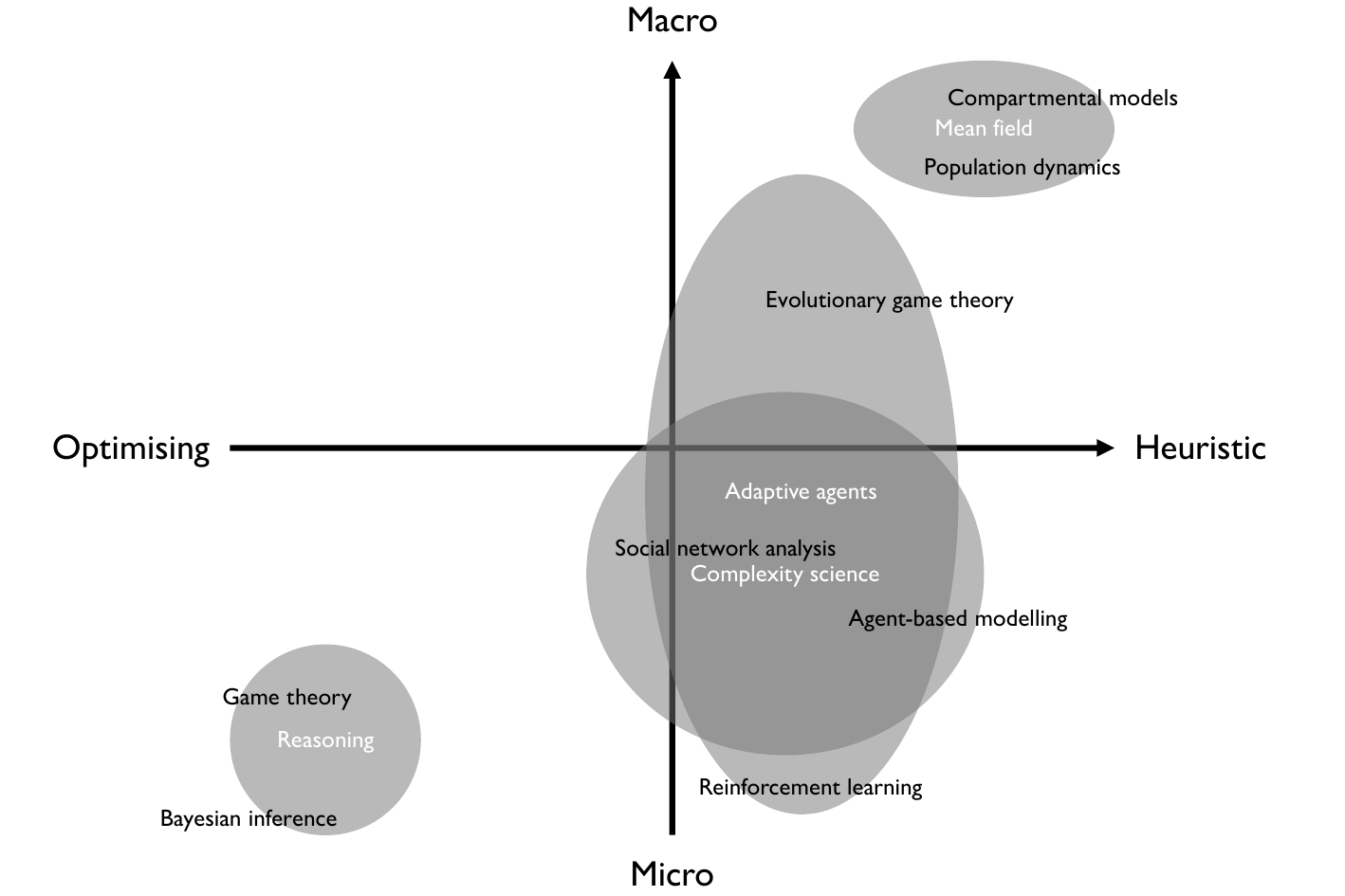}
\caption{Common modelling paradigms in cultural evolution. The paradigms are mapped out in a coordinate system according to agents’ level of evaluation and the scale of the modelled entities. The paradigms are grouped into four grey areas, representing groups of models.}
\label{fig5}
\end{figure}

We will here map out different modelling paradigms consistent with cultural evolution theory but that make different assumptions on actors' cognition, from the rational to the rule-based, and where the components and targeted phenomena operate at different levels, from the microlevel, focusing on individuals, to the macrolevel, focusing on populations. If we let these two dimensions be orthogonal axes, then we can place the different paradigms in a coordinate system (see Figure~\ref{fig5}).

Let us place optimising to heuristic on the first axis and micro to macro on the second. We then find models of reasoning in the lower left corner, since they target individual rationality. Here we find the paradigms of Bayesian inference and game theory, where rationality entails using information to build a sensible worldview or maximising outcomes, respectively. In the middle of the first axis and spanning the full range from micro to macro, we find models of adaptive agents. This includes reinforcement learning, which combines the previous two by being a microlevel model of using information for maximising rewards, but does so in a more rule-based manner. A more populational approach is evolutionary game theory, which considers strategic situations like classical game theory, but assumes selection pressures instead of rationality. In the upper right corner, we find mass-action or mean-field models of population dynamics or epidemiology, considering changes of frequencies in a population and spreads of diseases, respectively, when individuals are homogeneous and well-mixed. Finally, breaking the line from rational microlevel to rule-based macrolevel models, complexity science involves models that are typically focused on individual rule-based or adaptive actions and how they lead to emergent populational outcomes that are not simple aggregations of individuals actions. We will take a closer look at social network analysis and agent-based modelling, which focus on the structure of social interactions and autonomous heterogeneous agents, respectively. In the remainder of this chapter, we will walk through the different paradigms along with a few examples and finally map out a common template for modelling cultural evolution.

\subsection{Reasoning}\label{reasoning}

Some models focus on individual thinking in interaction with the environment or other agents, assuming that agents are either epistemically rational, as in Bayesian inference, or strategically rational, as in game theory.

\subsubsection{Bayesian inference}\label{bayesian-inference}

Bayesian models of cultural evolution formalise belief updating in response to evidence, and have been used for example to study polarisation of beliefs \citep{nielsen2021persistent,weatherall2021endogenous} and the emergence of linguistic structure \citep{griffiths2007language,kirby2014iterated}. They assume that agents are epistemically rational and use probabilistic reasoning based on all the available evidence. Agents start with initial beliefs, which form their priors, and which can differ based on cultural background or individual learning histories. Suppose that there is a hypothesis \(H\) that you are interested in. Your prior belief is a subjective probability \(P(H)\) for whether \(H\) is true. You then observe some evidence \(E\), supporting or working against \(H\). Being epistemically rational, this observation leads you to update your belief in \(H\) to the conditional probability \(P(H|E)\), the probability that \(H\) holds given that you have seen \(E\). The hypotheses can be about, for example, which scientific theory or treatment is better, grammar parameters in a language, which norm applies or if a coin is biased. The evidence could then be experimental results, language usage, observed behaviour or a sequence of coin flips. Again, since you are rational, for this you use Bayes' theorem, which says that

\[P(H|E) = \frac{P(E|H)P(H)}{P(E)}\]

As we can see, this also requires that you can estimate the likelihood \(P(E|H)\) to observe the evidence in the first place assuming that the hypothesis \(H\) is true. Sometimes this is a purely mathematical exercise (see e.g. the Monty Hall problem), sometimes it requires knowledge about the world. As an example of the latter, let us assume that you took a painkiller pill, but forgot to check its colour. You know it is either red or blue, but since you think there were about as many of each, they are equally likely to have been picked. Now you are concerned whether you are in the real world, \(H_{R}\), or in a matrix dream, \(H_{B}\), which are the somewhat unusual side effects of the respective pill, and your prior belief is \(P(H_{R}) = P(H_{B}) = \frac{1}{2}\). However, you realise that you still experience pain, which is likely in the real world, say \(P(E|H_{R}) = .7\), but unlikely in the matrix, say \(P(E|H_{B}) = .1\), so the probability of that experience is \(P(E) = P(E|H_{R})P(H_{R}) + P(E|H_{B})P(H_{B}) = .4\) (the probability that you had the red pill and then experience pain plus the probability that you had the blue pill and then experience pain). Given this new evidence,

\[P(H_{R}|E) = \frac{P(E|H_{R})P(H_{R})}{P(E)} = \frac{0.7 \cdot 0.5}{0.4} = 0.84
\]

Since we are already pretty deep into epistemology, let us take some examples from that literature on how models may surprise us with nonintuitive outcomes. Nielsen and Stewart \citep{nielsen2021persistent} investigated whether Bayesian agents will necessarily converge in their beliefs. Despite agents updating their beliefs rationally and receiving the same evidence, the authors showed that differences in prior beliefs can lead to divergent posterior beliefs, and under certain conditions (e.g. when assigning zero probability to some event others consider possible), these differences can persist or even widen over time. Thus, even ideally rational agents with access to the same evidence can experience persistent disagreement and polarisation. Relatedly, O'Connor and Weatherall found that such polarisation is possible even when one of the beliefs is false, if you treat other agents' evidence as uncertain \citep{oconnor2018scientific} and that this polarisation can spread to other beliefs as well \citep{weatherall2021endogenous}.

\subsubsection{Game theory}\label{game-theory}

Game theory may have been the first major formal framework with explicit ambitions to formalise the social sciences in a unified way, ``in the manner which has resulted in the establishment of physical science'' if possible \citep{vonneumann1944theory}. Contrasting to earlier efforts, modelling populations or individuals in isolation, game theory incorporates interdependent decision-making, which is arguably necessary to understand social behaviour. A game is thus a situation with given rules defining a set of strategies where the outcome for a player depends not only on their own decision but also on the decisions of the other players. The outcomes are in turn defined by the agents' preferences, or utilities.

A famous example is the \emph{Prisoners' Dilemma} (PD). Instead of prisoners, let us consider two companies competing for the same customers. Suppose that they provide the same goods for the same price, with an equal market share. If they are cooperative (towards each other, not the customers), then they will maintain their current prices and keep splitting the market between them. However, both of them have an incentive to lower their prices, or perhaps launch a costly advertising campaign, because then they can increase their market share. At the same time, this comes at the cost of a smaller revenue per sold item. If we stylise this situation and lump the different competitive strategies into one, called defect, and assign some values to the pairwise choices, then we can write the game as a two-by-two matrix with pairs of numbers representing the utility for the respective player. Suppose that the market is worth four zillions. Defecting costs a zillion, but attracts all customers of the competitor. These numbers may come across as a bit made-up, but for our analysis, it is only the relative utilities that matter, not the absolute numbers. Then we get the following matrix:

\begin{center}
\begin{tabular}{l|cc}
\diagbox[linecolor=white,linewidth=0pt]{Player 1}{Player 2} & Cooperate & Defect \\
\hline
Cooperate & \diagbox[linecolor=white,linewidth=0pt]{2}{2} & \diagbox[linecolor=white]{0}{3} \\
Defect    & \diagbox[linecolor=white,linewidth=0pt]{3}{0} & \diagbox[linecolor=white]{1}{1} \\
\end{tabular}
\end{center}


From the matrix we can see that mutual cooperation gives the highest total payoff, though one player can score higher by unilaterally defecting. The only strategy that is inferior in being (Pareto) dominated by another strategy is mutual defection, which is worse for both of them than mutual defection. Now to the interesting part: what will the players do?

We need to make some assumptions on how agents make decisions. With Bayesian inference above, they made the most out of the available information. Here we already have full information, but instead different preferences for the outcomes. In analogy to epistemic rationality, we assume economic rationality, where agents will maximise their expected payoff. This kind of pure self-interest may sound cynical, but the payoff is not necessarily monetary, but more generally utility, and then it becomes more or less tautological: agents maximise what they want (which may include well-being for others). The next part is trickier: not only do we assume that agents recognise the situation and know what they want, but they also know what the other player wants. You are rational, you assume your co-player is rational, and that they assume that you are rational, and so on. This is probably a key reason why economics is the social science where game theory is used extensively, applied to impersonal economic situations with monetary payoffs.

We could, however, also imagine more everyday situations. You and your flatmate might both have several opportunities to take out the rubbish, to the benefit of both of you, though you would like to avoid the hassle and hope that your flatmate does it, while running the risk it may remain stinking in the kitchen. This situation could also be modelled as a PD with a matrix similar to the above.

So, what would a rational actor do if the other player cooperates? Well, then cooperation gives a payoff of 2 and defection 3, so they would defect. If the other player defects, then cooperation pays 0 and defection 1, so again they would defect. Thus, no matter what the other player does, defection is the rational response. Since the game is symmetrical, the other player is in the same situation, so considering the case where the other player cooperates was purely hypothetical; they will both defect, be it competing with lower prices, annoying the general public with adverts or getting a filthy kitchen. Mutual defection is the only \emph{equilibrium} of the game: a pair of strategies where neither player has an interest in unilaterally changing their strategy. This is why the situation is a dilemma: we end up with a bad outcome even though there is an outcome, mutual cooperation, that we both agree would be better. Generalising to more than two players, we get a tragedy of the commons \citep[see e.g.][]{ostrom1990governing}, where we end up overusing valuable resources, since there is a benefit to us while the marginal cost is taken by society. Some examples are pollution, overgrazing and overfishing. A similar model could be used to explain why the lunchroom tends to be in a mess.

Not all is lost, though. We often play games repeatedly and there is a shadow of the future. My actions now may influence your actions tomorrow, or even someone else's actions, through rumours and reputation. If my cooperation leads to being met by cooperation, then it can become a rational choice. Mutual cooperation may also be achieved if we manage to maintain social norms where defection is disincentivised \citep[see][for an in-depth discussion]{jansson2026emergence}. These processes can turn the situation into a \emph{stag hunt} or some other \emph{coordination game}.

A simple example of a coordination game is what side of the road to drive on. Contrasting to the PD, this is clearly a situation with two equilibria: it does not matter so much which side we choose as long as we coordinate, and then we have no incentive to unilaterally change our strategy (with a potentially fatal outcome).

The stag hunt is an interesting special case, still with two equilibria, but with different qualities. Suppose that there is an upcoming party that you are not particularly interested in, but you suspect that a special someone (whom you are too intimidated to ask directly) might attend. It so happens that this special someone experiences the same considerations about you and the party. We could model this situation, where you both choose between going to the party and the comfortable option of staying at home, through the following matrix (with utilities rigorously expressed in emojis):

\begin{center}
\begin{tabular}{l|cc}
\diagbox[linecolor=white,linewidth=0pt]{Ego}{Alter} & Party & Home \\
\hline
Party & \diagbox[linecolor=white,linewidth=0pt]{\Large\Smiley}{\Large\Smiley} & \diagbox[linecolor=white,linewidth=0pt]{\Large\Sadey}{\Large\Neutrey} \\
Home  & \diagbox[linecolor=white,linewidth=0pt]{\Large\Neutrey}{\Large\Sadey} & \diagbox[linecolor=white,linewidth=0pt]{\Large\Neutrey}{\Large\Neutrey} \\
\end{tabular}
\end{center}

Like in the PD, there is a mutually desirable option: going to the party. Contrasting to the PD, this is an equilibrium: you would stay with it if Player 2 were there. Thus, for rational players with full information, this is an uninteresting game. If one of these assumptions is violated, however, as in this case (not necessarily assuming you are irrational, only lacking information), you risk sitting through a tedious party on your own. While both going to the party is the optimal equilibrium, staying at home is a risk-dominant equilibrium: no matter what the other person does, you know for sure that you will be having an okay time (perhaps reading this book) at home.

Game theory offers a rigorous way to analyse strategic interdependence, revealing the logic behind cooperation, conflict, and coordination in settings from market competition to household chores. However, this analysis depends on strong assumptions: that agents are rational, that they recognise the structure of the game, and that they form correct beliefs about others' rationality and preferences. These assumptions have been widely discussed and challenged, leading to extensions such as behavioural game theory, which incorporates insights from psychology, and evolutionary game theory, which explores how strategies spread in populations without requiring explicit reasoning. Even so, the standard models remain valuable as a clear baseline, which is idealised, but can itself often be surprisingly illuminating.

\subsection{Adaptivity}\label{adaptivity}

Another way to keep models simple and tractable is to make the opposite assumption: agents do not think at all. If they still respond to their environment by changing their strategies, then agents are adaptive. If we stay at the individual level, reinforcement learning studies how the individual interacts over time with its environment. If we rather think of the environment as other actors with strategies, then we are back at the interdependencies of game theory. However, void of rationality, we can instead assume a selective evolutionary pressure, at the population level. Not surprisingly, that paradigm is called evolutionary game theory. It is also possible to have multiple reinforcement learners, and under certain conditions there is an overlap between the two paradigms \citep{bloembergen2015evolutionary}.

\subsubsection{Reinforcement learning}\label{reinforcement-learning}

Reinforcement learning targets how agents, be it animals, humans or machines, learn to adapt their behaviour through experience. It is characterised by trial-and-error learning, where an agent interacts with its environment, observes outcomes in the form of rewards or punishments, and adjusts its actions accordingly. A parameter of interest is that there is a trade-off between exploration and exploitation. Going back to the very first example of choosing a restaurant, suppose you are a local with plenty of opportunities to eat out, that you ignore social information, and that you would happily become a regular at a good restaurant. How many restaurants should you explore before you exploit, that is, settle on one? And how many opportunities should you give each restaurant? Even if you have found a decent restaurant, there may still be a better one around. However, by trying a new restaurant, you run the risk of disappointment. And maybe you already found the best restaurant in town, but went there on a bad day or were unlucky with the specific dish you ordered.

Let us start with how you evaluate the information you received through your restaurant visit. One of the simplest and most influential models of learning is the Rescorla and Wagner \citep{rescorla1972theory} model, which formalises how associations between stimuli and outcomes are updated. This model was originally developed to explain Pavlovian conditioning in animals, and has since been broadly applied in psychology and neuroscience to describe how expectations are updated based on surprise or prediction error. In its simplest form, it can be described as:

\[\Delta V = \alpha(\lambda - V)\]

where \(V\) is the current associative strength, \(\lambda\) is the maximum strength a stimulus can acquire, and \(\alpha\) is the learning rate. Perhaps you associate Pizzeria Rodentio with excellent food experiences (\(V\)) and then in your next visit, you find a rat's tail in your pizza, which has a potentially highly negative value (\(\lambda\)), leading you to update the associative value proportionate to the discrepancy to your expectations and to, say, how attentive or sensitive (\(\alpha\)) you are, for example to rats as ingredients.

The question now is how the associative strengths guide your behaviour. If you have a specific number of options, then they should translate into probabilities for making each choice. A common model for this is the \emph{softmax} function. It balances exploration and exploitation, is indeed a probability function and is mathematically convenient. The balance is controlled by a `temperature' parameter \(\beta\). If an agent has estimated values \(V_{1},\ldots,V_{n}\) for the available choices \(A_{1},\ldots,A_{n}\), then the probability for choosing \(A_{i}\) is:

\[P(A_{i}) = \frac{e^{\beta V_{i}}}{e^{\beta V_{1}} + \ldots + e^{\beta V_{n}}}\]

If \(\beta = 0\), then the agent always explores, that is, associative values do not matter. The larger the \(\beta\), the more the agent exploits, with the risk of missing out on better options.

This model could be combined with social learning, for example as in the introductory example, to study trade-offs between both individual and social learning, and exploration and exploitation. Typically, a large amount of social learning is beneficial as long as there is a fair amount of individual learning and exploration in the population \citep[see e.g.][]{mesoudi2021simulation}.

Reinforcement learning is not only a paradigm for understanding learning and decision-making \citep{enquist2023human}, but as one of the basic machine learning paradigms, it has been successfully used to implement and develop artificial intelligence \citep{sutton2018reinforcement}, whether it is a robot navigating a maze or software mastering a game, a famous example being AlphaGo, the first compute program to beat a professional Go player, in 2015.

\subsubsection{Evolutionary game theory}\label{evolutionary-game-theory}

The logic of strategic interaction in game theory does not require agents to be consciously rational, only that they have a strategy. Recasting games as competitions among inherited (or socially transmitted) strategies rather than among forward-looking minds created the field of evolutionary game theory (EGT) \citep{smith1973logic,maynardsmith1982evolution}. EGT thus applies game theory to evolving populations. While classical game theory (CGT) has its roots in economics, EGT originates from biology, but has since then also been used in the social sciences, as an alternative to rational choice.

The underlying game structure is the same as in CGT, so we can analyse the same games assuming either rational actors or evolving populations. However, since the assumptions about the agents are different, so are the situations they tend to be applied to.

A canonical example is the Hawk--Dove game, which models conflict over resources. Imagine two individuals competing for a resource with value \emph{V}. A Hawk escalates every contest, running the risk that the opponent also escalates and potentially wins, in which case the agent pays an injury cost \emph{C}. The payoff to a Hawk meeting another Hawk is thus variable, but at expectation it is \((V - C)/2\). A Dove retreats if faced by aggression, thus sharing the resource when meeting another Dove. The payoff matrix for an actor (ego) depending on its own strategy and that of the other player (alter) is

\begin{center}
\begin{tabular}{l|cc}
\diagbox[linecolor=white,linewidth=0pt]{Ego}{Alter} & Hawk & Dove \\
\hline
Hawk & \diagbox[linecolor=white,linewidth=0pt]{$\tfrac{V-C}{2}$}{$\tfrac{V-C}{2}$} & \diagbox[linecolor=white]{$V$}{0} \\
Dove    & \diagbox[linecolor=white,linewidth=0pt]{0}{$V$} & \diagbox[linecolor=white]{$\tfrac{V}{2}$}{$\tfrac{V}{2}$} \\
\end{tabular}
\end{center}


Let \(x\) be the frequency of Hawks and \(\pi_{H}\) and \(\pi_{D}\) be the average payoffs to Haws and Doves, respectively. Using \emph{replicator dynamics} we can find that the frequency of Hawks changes (\(\dot{x}\)) over time as \(\dot{x} = x(1 - x)(\pi_{H} - \pi_{D})\). Hawks thus increase in frequency as long as there are both Hawks and Doves in the population and their average payoff is greater than that of Doves. We find an equilibrium, where frequencies remain constant, at \(\dot{x} = 0\), which happens when \(x = \tfrac{V}{C}\), provided that the cost of losing is greater than the value of the resource. Thus, aggressive and conciliatory behaviours coexist and depend on the ratio between value and cost. A potential everyday example of a scarce resource is the bottleneck that occurs when lanes are merging or crossing in traffic and people either hawkishly claim their right of way or dovishly negotiate who goes first.

Let us go back to the example of the Prisoners' Dilemma (PD) in the classical game theory section. We saw that rational actors are in a dilemma, because they will lose out on better outcomes by mutually defecting, unless there are future benefits to gain from repeated interactions. This kind of repeated games seems apt to study from an evolutionary perspective. Indeed, the political scientist Robert Axelrod \citep{axelrod1981evolution,axelrod1984evolution} early on invited researchers to submit strategies to a computer tournament to play a round-robin iterated PD, where the payoffs represent fitness and agents with successful strategies breed in greater numbers than unsuccessful ones. While unconditional cooperators played well against each other, unconditional defectors could easily exploit and invade a population of unconditional cooperators. However, when unconditional defectors become more numerous, they have to play against themselves more often, thus faring relatively poorly. It seems that rather than these simple and extreme strategies, quite a sophisticated exploitative strategy would be needed. Quite surprisingly, the winner was instead the simple Tit-for-Tat (TFT): cooperate on the first move, then copy the partner's previous move. Apart from being simple and clear, TFT is thus nice (by never defecting first), retaliatory (by punishing defection immediately) and forgiving (by restoring cooperation through a single cooperative act). With this, it withstands exploitation by defectors, while cooperating with its close relatives, showing how reciprocity can evolve without foresight.

In evolutionary biology, EGT has offered parsimonious mechanistic explanations for otherwise puzzling traits. However, while modelling payoffs as fitness of strategies, rather than as individual utilities, is a strength for explaining biological evolution, it brings more of a challenge when it comes to cultural evolution. Cultural fitness is not well-defined \citep[see e.g.][for discussions on this]{jansson2026cultural}. It is context-dependent, multidimensional, mediated by cognition, and not necessarily connected to retention or breeding of individual agents. One way to interpret cultural fitness is perhaps through the lens of reinforcement learning. Agents using reinforcement learning update the propensity to play a strategy in proportion to experienced or observed rewards. When many such agents interact, their aggregate dynamics approximate a replicator equation in which expected reward rate plays the role of fitness \citep{bloembergen2015evolutionary}. In this case, it may be more accurate to think of fitness in terms of retention rather than transmission.

\subsection{Mass action or mean field}\label{mass-action-or-mean-field}

Moving even further up the ladder from microlevel to macrolevel models, we find models studying populations as a whole, ignoring individual heterogeneity. Since we do not consider individual idiosyncratic processes, we can think of these populations as rule-based. To make this work, we assume the law of mass action: that rates of interaction scale with the product of the relevant state variables; in simpler terms, populations are well-mixed -- or, in even simpler terms, the chance of two types of individuals meeting is proportional to how many of each there are -- and each unit experiences the average effect of the rest. These processes are typically modelled as ordinary differential equations: rules of how one quantity changes at each moment based on its current value and the values of related quantities. Sometimes, as with mean-field models in physics, these are simpler tractable versions of more complex high-dimensional models to approximate their behaviour. We will look at \emph{population dynamics}, modelling one or more interacting populations over time, and \emph{compartmental models}, where parts of the population move between discrete ``compartments''.

\subsubsection{Population dynamics}\label{population-dynamics}

Population dynamics has a long history in mathematical biology, beginning with the Malthusian growth model \citep{malthus1798essay}, stating that all life forms have a propensity for exponential growth. This type of models studies the size and composition of populations over time as dynamical systems. To put it simply, the number of individuals \(N_{t + 1}\) in a population at a time \(t + 1\) is

\[N_{t + 1} = N_{t} + B_{t} - D_{t} + I_{t} - E_{t}\]

where the terms following the number \(N_{t}\) of individuals at time \(t\) are the number of births, deaths, immigrants and emigrants between times \(t\) and \(t + 1\). In a closed system without migration, we get a simple rate of change as

\[N_{t + 1} - N_{t} = B_{t} - D_{t} = \left( \frac{B_{t}}{N_{t}} - \frac{D_{t}}{N_{t}} \right)N_{t} = RN_{t}\]

or, expressed in continuous time,

\[\frac{dN}{dt} = rN\]

where \(r\) is the intrinsic rate of natural increase and \(\frac{dN}{dt}\) the derivative, the rate of change of \(N\) with time. If we assume that there is a carrying capacity \(K\), beyond which the population cannot grow, for example because of resource constraints, and that the population grows inversely proportionally to how much capacity is left, then the equation is multiplied by a limiting factor:

\[\frac{dN}{dt} = rN\left( 1 - \frac{N}{K} \right)\]

where \(\frac{N}{K}\) is the amount of used resources and \(1 - \frac{N}{K}\) thus the remaining amount. While the first continuous time equation produces exponential growth, this added factor produces the familiar s-shaped (logistic) growth curves.

We are typically interested in the dynamics of interacting populations. The Lotka--Volterra predator--prey model that was mentioned earlier, in the section on modelling as a practice, is an example that extends the dynamics to two interacting species. Let \(x\) and \(y\) be the population density of prey versus predator, respectively. The prey has an intrinsic growth rate \(\alpha\) and declines with the density of predators by a predation rate \(\beta\). Predators increase with density of prey by a conversion efficiency rate \(\gamma\) and decline naturally at a death rate \(\delta\). Since the rates are proportionate to the present size of the populations, we multiply the expression for prey by \(x\) and for predators by \(y\) and get the following differential equation system (\(\dot{x}\) is an alternative way to write the derivative, rate of change, of \(x\)):

\[\left\{ \begin{array}{rcl}
\dot{x} & =&  \alpha x - \beta xy \\
\dot{y} & = & \gamma xy - \delta y \\
\end{array} \right.\ \]

This relatively simple model generates oscillatory dynamics and counter-intuitive feedback (e.g. improving prey habitat can benefit predators more).

Modifications of this model have been used for example to study the dynamics of language decay \citep{abrams2003modelling}. Our species densities \(x\) and \(y\) are now frequencies of speakers of two languages, where, say, \(x\) refers to the present-day majority language. Let us ignore births and deaths and introduce a frequency bias \(\alpha\) similar to the introductory model on choosing a restaurant, so that the languages attract speakers at a frequency proportionate to \(x^{\alpha}\) and \(y^{\alpha}\), respectively. Apart from this, we will also introduce a status bias \(s\) for the majority language and a conversion ratio \(c\) to switch languages. The probability for a minority speaker to convert to the majority language is then \(csx^{\alpha}\) and vice versa \(c{(1 - s)y}^{\alpha}\). To derive the change in majority speakers over time, we multiply these expressions with the number of speakers that can convert from one to the other, that is, \(y\) and \(x\) respectively, and deduct the second term from the first. Let us first note that everyone in the population has one of these two languages as their primary language, so \(x + y = 1\) and thus \(y = 1 - x\). This means that we only need to count \(x\), resulting in one equation:

\[\dot{x} = csx^{\alpha}(1 - x) - c(1 - s)x(1 - x)^{\alpha}\]

With an appropriate choice of parameter values, this model provides a strong fit to historical data on speakers of Scottish Gaelic, Welsh and Quecha in the investigated societies. It shows that when a minority language lacks sufficient prestige or falls below a critical mass of speakers, then it faces an almost unavoidable slide towards extinction, unless its status is strengthened or conformity pressures are weakened.

\subsubsection{Compartmental models}\label{compartmental-models}

Instead of separate species or groups, compartmental models describe subgroups, or compartments, within the same population and individuals flowing between them. This type of models suddenly entered public awareness (and lent five minutes of fame to mathematical modellers) when the covid pandemic struck the world in 2020, since they are fundamental to epidemiology and the modelling of infectious diseases. A particularly featured subclass of models during the pandemic was the SIR models, where the letters indicate the compartments Susceptible, Infectious and Recovered. A timelapse of the model typically starts out with most agents being susceptible to a disease, except for a small number of agents who have already been infected and who are currently infectious, possibly a single patient zero. Susceptible agents become infectious at a contagion rate \(\beta\) when they meet infectious agents, and infectious agents spontaneously recover at rate \(\gamma\). If we denote the number or fractions of agents in the three compartments by \emph{S}, \emph{I} and \emph{R}, and assume that the population is well-mixed, so that infectious meet susceptible agents proportionate to rate \emph{$S\cdot I$}, then we get the differential equation system

\[\left\{ \begin{array}{rcl}
\dot{S} & = & - \beta SI \\
\dot{I} & = & \beta SI - \gamma I \\
\dot{R} & = & \gamma I \\
\end{array} \right.\ \]

Solving (or simulating) this system gives us curves representing the number of agents in the respective compartment over time (see the bottom left corner in Figure~\ref{fig3}). This is where the expression ``flatten the curve'' comes from: it was a key focus during the outbreak of the pandemic to keep \(\beta\) small (e.g. by maintaining social distancing) to make the peak of the curve of infected people as flat as possible, so that hospitals would not be overcrowded.

Not only pathogens are contagious, but so are ideas. Epidemiological models may be fruitful also for modelling cultural dispersal \citep{cavalli-sforza1981cultural}. The terminology may have to be interpreted metaphorically (though there are certainly some ideas and fads out there that we may be keener to label as diseases than others), but the equations remain the same.

So, for example, we could model the biological pathogen with a conventional SIR framework. We can then couple that to a parallel SIR-like model that casts anti-vaccine sentiment as a cultural pathogen that spreads alongside or fuels the epidemiological one. Using such a framework, it has been demonstrated how disease contagion interacts with anti-vaccine sentiments and assorted social contacts, suggesting potential strategies to mitigate spread of the disease \citep{mehta2020modelling}.

Extending this logic further, we can model purely the spread of ideas. For example, in the case of scientific ideas, Bettencourt et~al. \citep{bettencourt2006power} used an SIR‐style framework to fit historical data on the uptake of Feynman diagrams by theoretical physicists. In their case, ``infection'' means first publishing or using a Feynman diagram; ``recovery'' means moving on to other methods or retiring. They found that, unlike a typical disease where contact‐rate dominates, the diagram's long‐lived ``infectious'' tenure (once you adopted it, you tended to keep using it for decades) and organised social channels (e.g. conferences and collaborations) were the main drivers of its rapid uptake.

This mirrors how epidemiological terminology has been used frequently in memetics, where cultural traits, or memes, are often compared to parasites \citep{dawkins1976selfish} or viruses \citep{dennett1995darwins,blackmore1999meme}. Perhaps not so surprisingly, then, epidemiological models have been used to characterise the spread and diversification of religions \citep{doebeli2010model}. They have also been extended to include spontaneous non-social ``infection'', such as individual discovery of information or rewards, to study behavioural phenomena. A prominent case is obesity: although no biological pathogen is involved, weight-gain behaviours propagate through social ties in an SIR-like manner \citep{hill2010infectious}.

This brings us back to medical conditions and thus closes the loop, illustrating how epidemiological models can be used also to study cultural transmission, which can in turn explain the kind of phenomena (i.e. spread of diseases) that such models were initially designed to explain.

\subsection{Complexity}\label{complexity}

Finally, a direction that connects the micro to the macro is \emph{complex systems}, which studies macrooutcomes bottom up from self-organising components that interact with each other. Complexity does not necessarily refer to being complicated or having many parts, but specifically to the nonlinear, interconnected, and emergent nature of the system's behaviour. A clean example of this is the prediction of motion of a few celestial bodies, like planetary orbits. In a system with just two objects, like a star and a planet, their gravitational interaction can be easily calculated using Newton's laws of motion and gravity. This leads to predictable elliptical orbits. Add a third body, and their gravitational pulls interact in nonlinear ways. The system becomes complex: tiny changes in position lead to large changes in orbits. These orbits are chaotic and unpredictable in the long term. You cannot solve the dynamical system exactly, only simulate it. Even if the outcome is complex, the mechanism is very simple: each body just follows Newton's law of gravity, being pulled by the others proportionally to their masses and inversely to the square of the distance. It is the interactions that make the system unpredictable.

A key concept is emergence: simple rules at one level give rise to new patterns, behaviours or structures at a higher level that are not obvious from the rules themselves. For example, particles interact through physical forces, forming atoms; atoms combine into molecules with chemical properties; molecules form self-organising, replicating structures that become living cells; cells build multicellular organisms with nervous systems; neural activity gives rise to thoughts and emotions; and from the interaction of thinking individuals, social systems emerge. Each layer introduces new kinds of states and interactions shaped by, but irreducible to, the layer below.

To study social systems, we often model individuals as adaptive agents: unlike celestial bodies, they respond to their environment, learn from experience, and adjust their behaviour over time. When many such agents interact, their adaptations influence one another, creating feedback loops and evolving collective patterns. This gives rise to \emph{complex adaptive systems}, where the system's structure and dynamics emerge from the ongoing adaptation of its parts. Social systems, such as markets, norms or cultural trends, can thus be seen as complex adaptive systems: shaped by individual decisions, but producing outcomes no single agent controls or predicts.

Already the example of choosing restaurants in the introduction included both interactions between agents and nonlinearity in popularity, because of the conformity bias. If, instead of a conformity bias, we introduce heterogeneity, that is, we allow agents to be different, then we can find other feedback processes that give rise to more punctuated nonlinearity, with sudden leaps in popularity. Suppose that agents are not more attracted to more popular restaurants, but that they do avoid those that seem too empty. If agents employ different thresholds, say 10\% occupancy for some, 20\% for others, and even higher for the rest of the agents, then once a restaurant reaches 10\% occupancy, it will attract all those agents who now find it sufficiently filled. When those agents have arrived, the occupancy has quite possibly passed 20\%, and the restaurant now attracts also the more picky agents with that threshold, further increasing the occupancy, and so on. Restaurants can now quickly jump from 9\% occupancy to being fully seated by having a single additional igniting guest. Similar processes have been argued to underlie social movements and protests, such as the Arab spring, where people engage when sufficiently many other people do, for example when it feels safe enough \citep[see e.g.][]{granovetter1978threshold,howard2011opening,gonzalez-bailon2011dynamics,centola2018experimental}.

There is a whole book chapter \citep{michaud2026complex} dedicated to complex systems in this volume, and another one \citep{jansson2026dynamics} dedicated to cultural systems, so we will not dwell on the topic here. We will, however, briefly highlight two common modelling paradigms that capture heterogeneity: agent-based modelling (ABM) allows variation in agent properties, while social network analysis (SNA) focuses on variation in interaction patterns, when agents connect preferentially with some others. ABMs often include spatial or network-based structure, but SNA specifically studies how the topology of interactions shapes system dynamics.

\subsubsection{Agent-based models}\label{agent-based-models}

Agent-based models (ABMs) are computational tools designed to simulate the actions and interactions of autonomous agents within a defined environment. In contrast to models based on aggregate equations, ABMs explicitly incorporate individual-level heterogeneity and decision-making, typically with the aim of studying emergent phenomena that result from micro-level dynamics. Each agent operates according to a set of defined rules or adaptive behaviours, interacting with other agents and the environment, which might capture complex, non-linear processes inherent in social systems and cultural evolution. While ABMs are highly flexible in modelling detailed, realistic scenarios with nuanced interactions, this flexibility comes at the cost of reduced transparency and analytical tractability, and they are typically implemented as computer simulations.

While ABMs are often more complicated than other models discussed here, they are not necessarily more difficult to design; on the contrary, entry-level tools such as Netlogo make it easy for inexperienced learners of formal modelling to get started. Since it is easy to add parameters and additional assumptions, with many degrees of freedom, it is easy to find patterns that can be ascribed appealing narratives -- sometimes perhaps too easy. It is also easy to add too much and get lost in a jungle of parameters, where assumptions are too specific to be mapped to general human drives, while too generic to be mapped to empirical data, and where it is hard to disentangle what generates the outcome. The complexity and stochastic nature of agent-based simulations can make it challenging to clearly understand or predict system-level outcomes, and interpreting the results often requires extensive sensitivity analyses and computational experimentation. With many free parameters, it may be easy to miss that the ones driving the result are not actually the parameters of interest \citep{jansson2013pitfalls}.

As always, we should strive for as much simplicity as possible. Indeed, some of the most seminal models are simple enough for us to gain an intuition of what is going on, and to some extent even to be carried out using only pen and paper.

A clear example is Schelling’s segregation model \citep{schelling1971dynamic,schelling1978micromotives}. Imagine two types of agents living on a lattice. Each is content to be in the minority, as long as some neighbours are from their own group. For instance, suppose each agent requires at least one third of their neighbours to be of the same type; otherwise, they move to a random empty spot. If we begin with a random distribution and run simulations based on these rules, the result is striking: agents self-organise into large, homogeneous clusters. In other words, significant segregation emerges even when individuals are tolerant of diversity. This outcome can be understood through the concept of tipping. Even small local imbalances can trigger movement: if a few neighbours move out, then the proportion of same-group neighbours may drop below the threshold for others, prompting them to leave as well. Such local cascades can grow, reinforcing patterns of segregation. What begins as a mild preference for some similarity thus leads, step by step, to large-scale separation.

A related model, more in the context of cultural evolution, is Axelrod’s dissemination of culture model \citep{axelrod1997dissemination}. Agents still live on a lattice, but instead of moving, they interact with their immediate neighbours, exchanging cultural traits based on similarity. Each agent is characterised by a set of cultural features, each of which can adopt various trait values. Interaction occurs more frequently between agents sharing similar traits, leading them to become even more alike. Interestingly, this local convergence can generate global polarisation: although interactions drive cultural homogenisation locally, distinct cultural regions emerge at the macrolevel due to clusters forming with different trait combinations. Again, local and global patterns differ, one moving towards similarity, the other polarisation.

More complicated models can also provide new insights. We encountered an ABM already in the section on evolutionary game theory, where Axelrod (\citeyear{axelrod1984evolution}) conducted a simulation tournament of the prisoners' dilemma, resulting in the success of one of the simplest strategies. However, even though the simulations included many components, it is possible to understand what made the winning strategy successful.

\subsubsection{Social network analysis}\label{social-network-analysis}

Our final paradigm studies the topology of interactions between agents and what kind of dynamics this leads to. Social network analysis (SNA) focuses on how the structure of social ties (i.e., who interacts with whom) shapes the transmission and evolution of cultural traits. Rather than assuming that agents are homogeneously mixed or interact randomly, SNA highlights how properties of how nodes are connected influence cultural dynamics. Agents are represented as nodes and their connections as links in a network (a.k.a. graph), capturing patterns of social influence, communication or imitation. Some typical measures to consider are network size, number of connections (node degrees) and their distribution, centrality, and clustering \citep{newman2003structure,jackson2008social}.

Different social behaviours give rise to different types of networks: for instance, preferential attachment, where individuals are more likely to form ties with already well-connected others, can generate scale-free networks, characterised by highly connected hubs \citep{barabasi1999emergence}. Such networks, paired with preferential attachment, can generate rich-get-richer phenomena, as is the case for example for how new websites are more likely to link to already popular websites, collaboration networks, and the distribution of city sizes, where the population is inversely proportional to its rank within a country or region. In contrast, networks based on local interaction or social reinforcement often form small-world networks with high clustering and short average path lengths \citep{watts1998collective}. This entails and encapsulates the famous idea that all people are six or fewer social connections away from each other, commonly referred to as six degrees of separation, and it provides a mechanism for when and how such networks emerge. These structural differences have profound implications for how cultural traits spread. For example, simple contagions, such as viral videos or rumours, can rapidly diffuse through small-world networks. However, complex contagions, that is, behaviours requiring reinforcement from multiple peers, such as adopting a new social norm or technology, are more likely to spread in clustered networks where individuals are embedded in tightly-knit communities \citep{centola2007complex}.

A peculiar observation that can be understood by studying the topology of interactions is the so called \emph{friendship paradox}: most of the readers may have noted that their friends tend to have more friends than they do -- something that is particularly evident on social media, where connections are counted and displayed. ``Most readers?'' you may ask. Yes, and this does not (necessarily) indicate poor social skills, because on average, people's friends have more friends than they do. This phenomenon may sound paradoxical at first, but it emerges because people with more friends are more likely to be included in the friend lists of others, leading to an over-representation of highly connected individuals when considering the friends of a randomly selected person. This property has direct implications for cultural evolution. Information or behaviour observed in one's immediate network may disproportionately reflect the actions of popular individuals. Because of this, it can for example amplify perceived norms: if influential people adopt a practice, then it appears more common than it actually is, accelerating its diffusion \citep{lerman2016majority}. This effect can be exploited for interventions; for example, targeting central individuals in health campaigns has been shown to increase overall uptake \citep{banerjee2019using}, and one way to find those individuals is simply to ask people to ask their friends.

Social network analysis highlights that the structure of social transmission may be as important as its rules. The same transmission bias, say, can lead to entirely different population-level outcomes depending on the network. For instance, a norm that spreads easily in a well-connected network may fail to take hold in a more fragmented one \citep{centola2010spread}; polarisation may depend on connectivity; and perceived prevalence may diverge from actual frequency due to structural asymmetries like the friendship paradox.

SNA is often combined with agent-based models, to include heterogeneity between agents acting within a given network, to study how localised interactions scale into emergent macro-level outcomes. Since networks are formally represented as graphs, the same framework can be used both in theoretical models and in the analysis of real-world data, such as observed friendships, collaborations or online connections. To support this, statistical methods have been developed to describe, model, and analyse networks, for example to measure how central or clustered a network is, or to estimate how likely certain network structures are to form. Such methods thus help connect formal models to empirical observations and keep them grounded in reality.

\section{A general template for cultural evolution}\label{a-general-template-for-cultural-evolution}

The different paradigms can shed light on different aspects of cultural dynamics. As in the case of the blind men and the elephant, we can target the same phenomena with different types of models and gain very different insights, all true but for different aspects. An idea spreads because friends have adopted it, it is epistemically or strategically rational, it increases fitness, or it feeds on or outcompetes another idea. All these factors may play a role, but where did it come from, why did the friends adopt it, how do I know it makes sense, why does it give me an advantage, and how does it interact with other ideas? Can we combine the different parts and connect the pieces into a comprehensive theory? If we zoom out and spot the fully-fledged elephant, then all the different parts will make even more sense.

In the natural sciences, evolutionary theory has provided such an integrative framework. It unifies disparate observations, from the beak shapes of finches to the genetic structure of populations, into a coherent understanding of how biological diversity arises and persists. Evolutionary theory does not replace detailed models of, say, metabolic pathways or ecological interactions, but connects them, placing them within a common explanatory structure.

Cultural evolution provides a framework for linking the diverse mechanisms by which culture changes: the generation of variation in beliefs, practices and artefacts; the transmission of these variants through teaching, imitation, or communication; the differential success or influence of some variants over others; the retention of cultural traits in memory, institutions or traditions; and the transformation of ideas as they are interpreted, distorted or recombined. It does not seek to replace models of reasoning, learning, influence, or innovation, but rather to connect them, showing how diverse cognitive processes, social structures, and transmission mechanisms combine to produce and are produced by cultural change. When seen through this lens, phenomena as diverse as moral norms, linguistic patterns, technological diffusion or viral memes can be interpreted as parts of the same dynamic process, but with different particular properties.

To give a general template for modelling cultural evolution, we can take inspiration from Coleman's boat, as described in the introduction.

First we define the state of the system, involving the components for which we want to find a causal mechanism of change. This may be an aggregation of trait frequencies $F(t)$ at time $t$, such as the share of the population speaking a minority language, believing a particular story or practicing a certain behaviour. It may also be something that is not an aggregation of traits in the population, an institutional setting $I(t)$, such as norms, laws and formal organisations that structure behaviour. Agents are connected, with information flowing between them, through some kind of social structure $S(t)$, for example social networks. All these states are potentially endogenous, that is, they can be socially constructed within the system. We may also want to add exogenous influence from the environment, $E(t)$, such as resource scarcity or climate.

Going from the macro to the micro, trait frequencies, institutional rules, social structure and environmental cues and restrictions jointly shape what each individual perceives. This can be determined by how common a trait is, while the network filters whom you meet and skews the perceived distribution, and institutions and environment limit what you can do or see.

Each agent interprets what they were exposed to and combines it with internal rules and parameters, perhaps in view of current beliefs through Bayesian inference or some predefined adaptive rules. Rules might also be assumed to be innate. There may be some selection-like copying, for example with a conformity bias as in the restaurant example, or biased transformation towards cultural attractors \citep{sperber1996explaining}. In brief, exposure to the state of the world leads to mind-level updates.

Each mind-level update, in turn, leads to a behavioural choice, to produce an act: adopt, discard, tweak, trigger, update, cooperate or defect. A game theory model would consider this a strategic choice, while reinforcement learning relies on previous associations to the exposure. In other models, beliefs are revised, or the predefined and/or adaptive rules are executed. The behavioural response is guided and restricted by the institutional setting $I(t)$ and the environment $E(t)$.

The choice leads to a visible behaviour, for example to tell the story, wear the garment or use the tool. That behaviour becomes fresh evidence for everyone observing you. Who these observers are depends on the social structure and interaction topology S(\emph{t}), such as social networks. Again, each agent interprets what they were exposed to, combining it with internal rules and parameters, similar to above, but now observing an individual rather than a population-level phenomenon. This leads to mind-level updates and more new individual choices, and ultimately to a feedback loop of individual interactions.

As individuals start to behave differently, collective outcomes and the population picture shift. Cultural selection may have benefited variants linked to higher payoffs, prestige, conformity or other factors, or repeated reconstruction may have transformed narratives and practices towards psychological or ecological attractors. We have a new distribution of frequencies $F(t+1)$. The aggregation of changed behaviours may also lead to emergent phenomena, for example through threshold effects. This could for example bring about a new institutional setting $I(t+1)$. A straightforward example is when agents simply vote for a new policy. Social interactions may have been rewired, such as through technological innovations or followers flocking to a viral celebrity, causing a new social structure $S(t+1)$. The environment, $E(t+1)$, may also have changed, by itself or as a consequence of cultural practices, for example through overharvesting. We now have a new state of the system, and all the steps are repeated under these new conditions.

As we have seen, not all models bring detail to all steps, but tend to focus on some of them. Together we get the full picture. Mass-action or mean-field models tend to draw the arrow directly from $[F(t), I(t), S(t), E(t)]$ to $[F(t+1), I(t+1), S(t+1), E(t+1)]$ without really dipping into the causal mechanisms. Complex systems models provide the link from the micro to the macro and to some extent from the macro to the micro, along with adaptive models. Many models of cultural evolution focus on the feedback loop at the microlevel, where individuals transmit cultural traits to other individuals. From the lens of individuals, we can use an even simpler template:

Agents have traits (variation), most of which are cultural and transmissible. By interacting (transmission), agents learn about the traits of others. Agents transmit and pick up traits (selection) based on properties of the environment, the population, the sender and the receiver, and the trait itself. Agents then have a new distribution of traits, and the loop continues. Over time, we get an evolutionary process on cultural traits.

This template assumes a lot of the action in selective step, and the rules are often assumed as a black box that just works, perhaps supported by empirical observations. Just as individual-level interactions give rise to emergent societal phenomena, it is possible that the individual-level selective mechanisms are emergent, from the interactions between cultural traits, and how they interact with individuals and the environment \citep{buskell2019systems,jansson2021modelling,jansson2023cultural}. We look into the role of such cultural systems in the final chapter of this volume \citep{jansson2026dynamics}.

Formal modelling is a powerful and versatile tool for understanding cultural evolution. It provides clarity and precision, showing how simple assumptions can generate complex societal patterns. The critical question is not whether to build simplified models of reality -- indeed, all researchers, and for that matter the average Joe, have their mental models -- but whether we make these models explicit and open for exploration, scrutiny and critique. Clearly articulated models do not necessarily narrow our view; instead, they guide us to think more systematically and deeply.

General theories and minimalist, generic models serve as important foundations, offering conceptual frameworks from which more detailed explanations can develop. The surest way to tame complexity may be to start with transparent ``sample models'' that capture only the core causal forces \citep{boyd1985culture}. By focusing on fundamental processes, these basic models reveal essential mechanisms, clarify how individual actions scale up to societal outcomes, and allow us to test qualitative predictions before adding additional layers of complexity. Thus, a core set of simple models helps to organise and inform more elaborate and specialised theories with predictive power.

Throughout this chapter, we encountered evolutionary dynamics in diverse forms: skewed popularity driven by conformity biases; persistent polarisation among rational agents; cycles of cooperation and defection in adaptive games; gradual shifts in language dominance; and spontaneous segregation arising from simple individual preferences. Despite their variety, these models share a common evolutionary logic based on replication, variation, and selection.

The value of formal modelling goes beyond just explaining specific phenomena. Translating intuitive ideas into explicit mathematical structures cultivates precise analytic thinking, sharpens conceptual understanding, and promotes dialogue across disciplines. Proficiency in modelling improves with practice. For readers interested in deepening their understanding, exploring existing models or experimenting with their own can be especially rewarding. Whether your goal is to understand, predict, or explain cultural change, formal models offer robust tools for structured inquiry and clear thought.

\bibliographystyle{fredrik}
\bibliography{inpress,references}

\end{document}